%% file: main.tex
\documentclass[conference]{IEEEtran}
\IEEEoverridecommandlockouts
\usepackage[nolist]{acronym}
\usepackage{algorithm}
\usepackage[noend]{algpseudocode} 
\usepackage{amsmath,amssymb,amsfonts}
\usepackage{array}
\usepackage{caption}
\usepackage{cite}
\usepackage{eqparbox}

\usepackage{graphicx}
\usepackage{layouts}
\usepackage{macros}
\usepackage{pgfplots}
\pgfplotsset{compat=newest} 
\usepgfplotslibrary{statistics} 

\usepackage{siunitx}
\usepackage{subfig}
\usepackage{booktabs}

\usepackage{textcomp}

\usepackage{todonotes}
\usepackage{url}
\usepackage{xcolor}
\usepackage{xspace}

\usepackage{tikz}
\usepackage{tikzscale}
\usetikzlibrary{matrix}
\usetikzlibrary{arrows.meta, backgrounds, calc, fit, decorations.pathreplacing,decorations.pathmorphing, fit, positioning, shapes, shadows.blur, shapes.symbols}
\pgfdeclarelayer{foreground}
\pgfsetlayers{background,main,foreground}
\definecolor{darkblue_orig}{HTML}{000435}
\colorlet{darkblue}{darkblue_orig!70!white}
\definecolor{tudresden}{RGB}{0,48,93}
\colorlet{eNodeColor}{tudresden!50!white}
\definecolor{byzantium}{HTML}{702963}
\definecolor{burgundy}{HTML}{953553}
\definecolor{darkgreen_orig}{HTML}{184632}
\colorlet{darkgreen}{darkgreen_orig!90!white}
\definecolor{darkgray}{HTML}{575757}

\usepackage{balance}
\usepackage{stfloats}

\def\BibTeX{{\rm B\kern-.05em{\sc i\kern-.025em b}\kern-.08em
    T\kern-.1667em\lower.7ex\hbox{E}\kern-.125emX}}

\begin{document}

\title{\dez: Decentralized $z$-Anonymity with Privacy-Preserving Coordination
\thanks{This research has been funded by the Federal Ministry of Research, Technology and Space (BMFTR, Project~16KISA034) and the European Union (NextGenerationEU).}
}

\author{
    \IEEEauthorblockN{Carolin Brunn and Florian Tschorsch}
    \IEEEauthorblockA{\textit{Chair of Privacy \& Security} \\
    \textit{Dresden University of Technology (TU Dresden)}, Germany \\
    \texttt{\{carolin.brunn, florian.tschorsch\}@tu-dresden.de}}
}

\maketitle
\begin{acronym}
    \acro{ce}[CE]{central entity}
    \acro{ccc}[CCC]{clock cycle coordinator}
    \acro{hbc}[HBC]{honest-but-curious attacker}
    \acro{gw}[GW]{gateway}
    \acro{sn}[SN]{sensor node}
    \acro{cbf}[CBF]{counting bloom filter}
    \acro{cnt}[$cnt$]{counting structure}
    \acro{bss}[BSS]{Basic Secure Sum protocol}
    \acro{iot}[IoT]{Internet of Things}
\end{acronym}

\input{0_abstract}
\input{1_introduction}

%
\input{2_central_zA}
\input{3_sysMod}
\input{4_deZent_protocol}

\input{5_privInDez}

\input{6_methodology}
\input{7_evaluation}

\input{8_relatedWork}

\input{9_conclusion.tex}

\bibliographystyle{IEEEtran}
\bibliography{references}
\input{appendix.tex}
\end{document}

%% file: 0_abstract.tex
\begin{abstract}

Analyzing large volumes of sensor network data, such as electricity consumption measurements from smart meters, is essential for modern applications but raises significant privacy concerns.
Privacy-enhancing technologies like \zanon offer efficient anonymization for continuous data streams by suppressing rare values that could lead to re-identification, making it particularly suited for resource-constrained environments.
Originally designed for centralized architectures, \zanon assumes a trusted central entity.
In this paper, we introduce \dez, a decentralized implementation of \zanon that minimizes trust in the central entity by realizing local \zanon with lightweight coordination.
We develop \dez using a stochastic counting structure and secure sum to coordinate private anonymization across the network.
Our results show that \dez achieves comparable performance to centralized \zanon in terms of publication ratio, while reducing the communication overhead towards the central entity.
Thus, \dez presents a promising approach for enhancing privacy in sensor networks while preserving system efficiency.

\end{abstract}

%% file: 1_introduction.tex
\section{Introduction}

Analyzing large data streams from sensor networks is increasingly vital for both everyday applications and industrial processes, with demand expected to grow further.
In the \ac{iot}, sensors are an essential component used to monitor environmental parameters such as temperature or electricity consumption~\cite{Alshohoumi2019IoTArch,pallas2021RedCastle}.
Collecting environmental data, however, also raises privacy concerns: continuous measurements can reveal sensitive information and even identify which TV series is being watched~\cite{Carmody2021AIPriv,Finster2015PrivSMSurvey,Schirmer2020IdentificationOT}.

To protect client privacy in sensor networks, it is essential to implement privacy‑enhancing technologies.
Various strategies have been proposed for continuous data streams, including \kanon for data streams~\cite{Cao2011Castle,sweeney2002kAno,Samarati2001kAno} and differential privacy~\cite{dwork2010DP}, both of which have proven effective.
More recently, \zanon~\cite{Jha2020zAnon,Jha2023zAnon} has emerged as a lightweight and memory‑efficient approach that operates almost in real time by immediately deciding whether to publish or suppress each measurement.
It protects privacy by withholding rare values that could enable re‑identification~\cite{Jha2023zAnon}.
In addition to its suggested applications in transaction data, network traffic, and location trajectories~\cite{Jha2020zAnon,Sedlak22zUse}, we see significant potential for \zanon in \ac{iot} networks in general, where large volumes of high‑frequency data are generated.

\zanon was originally designed for centralized systems in which all data is collected at a single data sink.
However, given the distributed nature of sensor networks and the need to minimize trust in a \acl{ce} capable of extensive system-wide analysis, a decentralized solution is required.

In this paper, we propose \dez, a decentralized implementation of \zanon designed for sensor networks.
Our approach preserves data utility while reducing trust requirements and eliminating the need for a
coordinating \acl{ce}.
The main contribution of \dez is a privacy-preserving approach to distributed counting and data publication, combining secure summation with a stochastic counting structure.
This design enables \aclp{gw} to locally $z$‑anonymize data without leaking information, making \dez particularly suitable for the distributed and high‑frequency nature of sensor networks.

Our findings suggest that \dez produces results comparable to centralized \zanon by~\cite{Jha2023zAnon} based on metrics such as message count and publication ratio.
While decentralization requires additional coordination among the \aclp{gw}, it reduces the overall number of messages sent to the \acl{ce}.
Although the coordination process requires additional memory at the \aclp{gw}, this trade-off seems acceptable given that they are system entities with greater capabilities.
We thus conclude that \dez successfully decentralizes anonymization in distributed networks, effectively reducing trust.

The paper is organized as follows.
We first introduce \zanon as the core privacy technology for \dez
followed by our system and attacker model.
Sec.~\ref{sec:deZent} presents \dez before we describe the privacy preserving implementation.
In Sec.~\ref{sec:methodology}, we describe our methodology, while Sec.~\ref{sec:eval} presents our evaluation results.
Before concluding the paper, we discuss related work in Sec.~\ref{sec:relWork}.

%% file: 2_central_zA.tex
%%%%%%%
\section{\zanon}
\label{sec:cent_zanon}

Due to its low memory and computational overhead, we recognize \zanon as an effective strategy to enhance privacy in 
\ac{iot} networks.
\zanon was introduced in~\cite{Jha2020zAnon,Jha2023zAnon} and shared on GitHub.\footnote{\url{https://github.com/nikhiljha95/zanonymity/blob/main/README.md}}
\zanon enables anonymity without delays when publishing data streams while preventing re-identification based on rare sensitive attributes.

More precisely, \zanon processes one potentially identifying attribute, such as a sensor value.
\zanon operates on continuous data streams, processing each data point only once and publishing it immediately if at least $z$ clients have reported the same value within a specified time period, $\Delta t$.
This method effectively suppresses rare values from smaller groups of users and ensures an implicit anonymity set.
Each published data point also includes a timestamp and a client pseudonym.
The parameters $z$ and $\Delta t$ can be adjusted to balance data utility and privacy.
As shown in~\cite{Jha2023zAnon}, the output of \zanon is also \kous with a certain probability, which depends on various parameters, including $z$, $\Delta t$, and the ratio between number of users and $z$.
Generally, the higher the value of $z$, the greater the probability of \kous output.

At present, the implementations are confined to a centralized setting.
That is, \zanon could be applied at a \ac{ce} that can acquire significant insights about the system and its individual clients, as it can collect, and analyze all data of the system.
This raises trust concerns, as the \ac{ce} may draw accurate conclusions leading to the correlation of private information.
As a result, implementing \zanon on the \ac{ce} imposes high trust requirements towards the \ac{ce}.

%% file: 3_sysMod.tex
\section{System Model}
\label{sec:sysmodel}

We aim to reduce trust requirements towards a \ac{ce} and to enhance privacy by implementing anonymization at local components of the distributed system.
We focus on realizing \dez for \ac{iot} networks that may contain different sensors in a distributed setup.
An exemplary use case that uses such a system is smart metering of electricity consumption in district management, encompassing data from households, industry, and local businesses~\cite{Barai2015Arch}.
In the following, we describe the system model, including the distributed architecture and the capabilities of the system entities.

%%%%%%%
\subsection{System Architecture}
\label{subsec:arch}

\begin{figure}[t]
    \centering
    \subfloat[Without coordination.]{
        \begin{minipage}[b]{0.39\linewidth}
          \resizebox{\textwidth}{!}{\input{sys_arch_sink.tikz}\unskip}
          \label{fig:arch_wo}
        \end{minipage}
    }\hfill
    \subfloat[With coordination.]{ 
        \begin{minipage}[b]{.51\linewidth}
          \resizebox{\textwidth}{!}{\input{sys_arch.tikz}\unskip}
          \label{fig:arch_w_comm}
        \end{minipage}
    }%
    \caption{\footnotesize System architecture of a distributed sensor network.}
    \vspace{-15pt}
\end{figure}
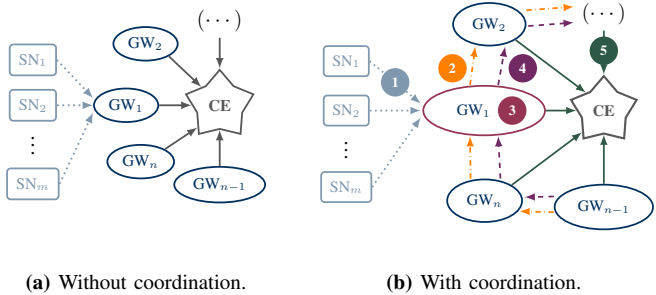

Sensor networks, such as smart metering infrastructures, primarily consist of distributed hierarchical systems, which include multiple logical layers with distinct processing and communication capabilities~\cite{Barai2015Arch,Bikmetiv2015Arch,Raza2023Arch}.
In Fig.~\ref{fig:arch_wo}, we illustrate a representative system architecture for smart metering ~\cite{Raza2023Arch}.

The \acp{sn} are installed on the client side and regularly measure data.
These \acp{sn} have mostly limited processing and communication capabilities.
Therefore, \acp{sn} are connected to a trusted \ac{gw}, which serves as an intermediary step to forward data from the \acp{sn} to the \ac{ce}, for example, an energy provider.
To minimize complexity for \acp{sn}, tasks like measurement requests and coordination can be initiated by \acp{gw}.
This further reduces power usage, for example through the use of wake-up receivers.
In the exemplary smart metering use case, power consumption is not a major concern, but it is crucial for self-sufficient sensor networks that lack regular power supply.

The \acp{gw} are robust system entities with greater computational and memory resources.
They collect data from several \acp{sn}, perform additional processing steps, and most importantly, forward the data to a \ac{ce}. 
The \acp{gw} can potentially use more powerful communication technologies, e.g., by implementing 5G connectivity, thereby enabling higher bandwidth and throughput.

The \ac{ce} serves as a data sink, centrally collecting all metered data from the system.
By receiving data points from the entire system at regular intervals, the \ac{ce} is capable of performing extensive analyses.
The anonymization step of centralized \zanon (cf. Sec.~\ref{sec:cent_zanon}) is executed on this \ac{ce}, which requires a certain degree of trust from the clients.

To reduce this trust requirement, we develop \dez to leverage the advantage of the distributed system and implement \zanon on the \acp{gw}, such that the \ac{ce} only receives anonymized data.
To this end, \dez assumes communication links between the \acp{gw}, which are organized in a ring topology as illustrated in Fig.~\ref{fig:arch_w_comm}.

\dez works with clock cycles, repeating the same steps per cycle and, therefore, requires semi-synchronized clocks.
The synchronization is needed to identify the current clock cycle during anonymization.
Consequently, the required precision correlates with the desired measurement frequency.
In the exemplary use case of electricity consumption metering, data are often measured every 15~minutes~\cite{Raza2023Arch}, which requires a precision of up to 7 minutes that can be easily realized.

%%%%%%%
\subsection{Attacker Model}
\label{subsec:attack_model}

We assume an \ac{hbc} who follows the protocol but attempts to infer sensitive information about clients.
While stronger attacker models certainly exist, an \ac{hbc} is reasonable in our use case, as smart energy metering operates in a highly regulated environment with controlled infrastructure.
Extending \dez to defend against stronger, active attackers would broaden its applicability, but would likely incur performance penalties.

Therefore, our main goal with \dez is to prevent re-identification and the inference of personal characteristics, e.g., linking consumption patterns to specific clients.
We focus on protecting against a corrupt \ac{ce} that can combine and analyze data from all \acp{sn}.
At the same time, we seek to deploy a solution that is lightweight and can be applied in sensor networks.
We further assume secure communication channels between entities, which excludes eavesdroppers from our threat model.
The adversary's inference capabilities are therefore limited to the data available through protocol-compliant interaction and published outputs, not through side channels such as traffic analysis.
We further assume that \acp{sn} trust their associated \ac{gw} and that widespread collusion among \acp{gw} is unlikely.
Under this assumption, \dez offers probabilistic protection against colluding attackers: since neighboring \acp{gw} are selected pseudo-randomly when establishing the ring topology, an attacker cannot reliably position colluding \acp{gw} next to a target.
We analyze the collusion probability in Sec.~\ref{subsec:privDez}.

%% file: sys_arch_sink.tikz
%% tikzstyles
\tikzstyle{ce_node}=[shape=star, font={\small}, draw=darkgray, text=darkgray, minimum height=0pt, line width=1pt, inner sep=3pt]%
\tikzstyle{m_node}=[shape=ellipse, font={\small}, draw=tudresden, text=tudresden, line width = 1pt, rounded corners=2pt, inner sep = 3pt]%
\tikzstyle{e_node}=[shape=rectangle, font={\small}, draw=eNodeColor, text=eNodeColor, line width = 1pt, rounded corners=2pt, inner sep = 5pt, font=\small]%

\tikzstyle{ce_m_conn}=[-latex, draw=darkgray, line width = 1pt]%
\tikzstyle{m_e_conn}=[-latex, draw=eNodeColor, style = dotted, line width = 1pt]%

%% tikz picture
\begin{tikzpicture}[scale=1, baseline=(current bounding box.south)]

    %%% CE
    \node[style=ce_node] (ce) {\bf CE};
    %%% m
    \begin{scope}[node distance=17pt and 18pt]
        \node[style=m_node, left=of ce] (m_1) {\acs{gw}$_1$};
        \node[style=m_node, above left=of ce] (m_2) {\acs{gw}$_2$};
        \node[draw=none, above =of ce] (dots_1) {\large $(\cdots)$};
        \node[style=m_node, below=20pt of ce] (m_n1) {\acs{gw}$_{n-1}$};
        \node[style=m_node, below left=of ce, yshift=5pt] (m_n) {\acs{gw}$_n$};
    \end{scope}
    %%% E
    \begin{scope}[node distance=7pt and 20pt]
        \node[style=e_node, left=of m_1] (e_2) {\acs{sn}$_2$};
        \node[style=e_node, above=of e_2] (e_1) {\acs{sn}$_1$};
        \node[below=-1pt of e_2] (e_dots) {\large $\vdots$};
        \node[style=e_node, below=4pt of e_dots] (e_m) {\acs{sn}$_m$};
    \end{scope}

    %%% m - CE connection
    \draw[style = ce_m_conn] (m_1.east) -- (ce.west);
    \draw[style = ce_m_conn] (m_2.south east) -- (ce.north west);
    \draw[style = ce_m_conn] (dots_1.south) -- (ce.north);
    \draw[style = ce_m_conn] (m_n1.north) -- (ce.south);
    \draw[style = ce_m_conn] (m_n.north east) -- (ce.south west);
    
    % m - E connections
    \draw[style=m_e_conn] (e_1.east) -- ([yshift=4pt]m_1.west);
    \draw[style=m_e_conn] (e_2.east) -- (m_1.west);
    \draw[style=m_e_conn] (e_m.east) -- ([yshift=-4pt]m_1.west);

    \node[below=8pt of m_n1] (DUMMY) {};
\end{tikzpicture}

%% file: sys_arch.tikz
%% tikzstyles
\tikzstyle{ce_node}=[shape=star, font={\small}, draw=darkgray, text=darkgray, minimum height=0pt, line width=1.2pt, inner sep=3pt]%
\tikzstyle{m_node}=[shape=ellipse, font={\small}, draw=tudresden, text=tudresden, line width = 1pt, rounded corners=2pt, inner sep = 5pt]%
\tikzstyle{e_node}=[shape=rectangle, draw=eNodeColor, text=eNodeColor, line width = 1pt, rounded corners=2pt, inner sep = 5pt, font=\small]%
\tikzstyle{label}=[shape = circle, line width = 0.8pt, font=\footnotesize\bfseries]

\tikzstyle{ce_m_conn}=[-latex, draw=darkgreen, line width = 1pt]%
\tikzstyle{m_m_conn_col}=[-latex, style = dashdotted, draw=orange, line width = 1pt]%
\tikzstyle{m_m_conn_pub}=[-latex, style = dashed, draw=byzantium, line width = 1pt]%
\tikzstyle{m_e_conn}=[-latex, draw=eNodeColor, style = dotted, line width = 1pt]%

%% tikz picture
\begin{tikzpicture}[scale=1, baseline=(current bounding box.south)]

    %%% CE
    \node[style=ce_node] (ce) {\bf CE};

    %%% M ring
    \begin{scope}[node distance=30pt and 40pt]
        \node[style=ellipse, font={\small}, text=tudresden, fill=white, left=of ce] (m_1) {\acs{gw}$_1$};
        %label 3
        \node[style=label, fill=burgundy, text=white, right=-5pt of m_1, align=left] (l3) {3};
        \node[ellipse, fit=(m_1)(l3), draw=burgundy, line width = 1pt, inner sep=0pt] (m_l_1) {};

        \node[style=m_node, above left=of ce] (m_2) {\acs{gw}$_2$};
        %dots
        \node[draw=none, above =25pt of ce] (dots_1) {\large $(\cdots)$};
        \node[style=m_node, below=28pt of ce] (m_n1) {\acs{gw}$_{n-1}$};
        \node[style=m_node, below left=of ce] (m_n) {\acs{gw}$_n$};
    \end{scope}

    %%% E
    \begin{scope}[node distance=10pt and 30pt]
        \node[style=e_node, left=of m_l_1] (e_2) {\acs{sn}$_2$};
        \node[style=e_node, above=of e_2] (e_1) {\acs{sn}$_1$};
        \node[below=-1pt of e_2] (e_dots) {\large $\vdots$};
        \node[style=e_node, below=4pt of e_dots] (e_m) {\acs{sn}$_m$};
    \end{scope}

    %%% M - CE connection
    \draw[style = ce_m_conn] (m_l_1.east) -- (ce.west);
    \draw[style = ce_m_conn] (m_2.south east) -- (ce.north west);
    \draw[style = ce_m_conn] (m_n1.north) -- (ce.south);
    \draw[style = ce_m_conn] (m_n.north east) -- (ce.south west);
    \draw[style = ce_m_conn] (dots_1.south) -- (ce.north);

    % M - M connections - collection
    \draw[style = m_m_conn_col] ([xshift=-8pt]m_l_1.north) -- ([xshift=-8pt]m_2.south);
    % label 2
    \node[style=label, fill=orange, text=white, above=1pt of m_l_1, xshift=-18pt] (l2) {2};
    \draw[style = m_m_conn_col] ([yshift=8pt]m_2.east) -- ([xshift=-1pt, yshift=4pt]dots_1.west);
    \draw[style = m_m_conn_col] ([yshift=-3pt]m_n1.west) -- ([xshift=-2pt, yshift=-6pt]m_n.east);
    \draw[style = m_m_conn_col] ([xshift=-8pt]m_n.north) -- ([xshift=-8pt]m_l_1.south);

    % M - M connections - publication
    \draw[style = m_m_conn_pub] ([xshift=8pt]m_l_1.north) -- ([xshift=9pt]m_2.south);
    %label 4
    \node[style=label, fill=byzantium, text=white, above=1pt of m_l_1, xshift=22pt] (l4) {4};

    \draw[style = m_m_conn_pub] ([xshift=1pt, yshift=-1pt]m_2.east) -- ([xshift=-1pt, yshift=-5pt]dots_1.west);
    \draw[style = m_m_conn_pub] ([yshift=6pt]m_n1.west) -- ([xshift=1pt, yshift=2pt]m_n.east);
    \draw[style = m_m_conn_pub] ([xshift=9pt]m_n.north) -- ([xshift=8pt]m_l_1.south);
    % M - E connections
    % label 1
    \draw[style=m_e_conn] (e_1.east) --
        node[style=label, fill=eNodeColor, text=white] (l1) {1}
        ([xshift=-1pt, yshift=4pt]m_l_1.west);
    \draw[style=m_e_conn] (e_2.east) -- ([xshift=-1pt]m_l_1.west);
    \draw[style=m_e_conn] (e_m.east) -- ([xshift=-1pt, yshift=-4pt]m_l_1.west);
    % label 5
    \node[style=label,  fill=darkgreen, text=white, below=3pt of dots_1] (l5) {5};
\end{tikzpicture}

%% file: 4_deZent_protocol.tex
\section{\dez - Local \zanon with Coordination}
\label{sec:deZent}

To avoid placing trust in the \ac{ce}, we aim to enforce \zanon at \acp{gw}, which we refer to as \emph{local \zanon}.
This approach enhances client privacy by limiting trust to local components and reducing global data visibility.
However, separate \acp{gw} can only process data of their respective \acp{sn} which might reduce utility or anonymity.
To address this, we develop the privacy-preserving coordination protocol \dez, which enables local \zanon with minimal coordination across \acp{gw}.
The coordination offered by \dez is used primarily to perform distributed counting among the \acp{gw} and to assign responsibility for publishing anonymized tuples.
Due to this coordination, data from all \acp{sn} within the system are considered during anonymization, thus ensuring a high anonymity and utility level.

\dez repeats the same steps per clock cycle.
In each cycle, one \ac{gw} is designated as \ac{ccc}.
To perform distributed counting and subsequent data publication, the \acp{gw} perform two communication rounds in a round-robin fashion.
The resulting data flow and coordination steps in these rounds can be summarized as follows~(step numbers correspond to those shown in Fig.~\ref{fig:arch_w_comm}):
\begin{enumerate}
    \item \acp{sn} measure data and send it to their associated \ac{gw}.
    \item In the collection round, the \acp{gw} exchange counts of each measurement along the ring.
    \item The \ac{ccc} (e.g., $\ac{gw}_1$ in Fig.~\ref{fig:arch_w_comm}) ensures that only values with at least $z$ occurrences are designated for publication by deleting lower values in the counting structure.
    \item In the publication round, the \acp{gw} pass the updated counting structure along the ring to designate publishing responsibilities.
    \item \acp{gw} responsible for publication forward the data to the \ac{ce}, which collects it for further processing.
\end{enumerate}

Through \dez, the \acp{gw} can collectively enforce the same $z$ threshold
as would be achieved if \zanon were executed directly on the \ac{ce}.

In the following, we provide a detailed description of our \dez protocol,
for achieving local \zanon with minimal coordination.
Accompanying pseudocode can be found in Appendix~\ref{app:dezCode}.
Additionally, we provide an implementation as part of our simulation.\footnote{\url{https://github.com/carolin-brunn/deZent-local_zanon}}
For now, we focus on the coordination algorithm; the privacy mechanisms are discussed in Sec.~\ref{sec:privInDez}.

%%%%%%%%%%%%%%%%%%%%%
\subsubsection{System Setup}

During setup, connections and ring topology between \acp{gw} are established autonomously.
At the beginning of each clock cycle, a \ac{ccc} is designated from the set of \acp{gw}. 
Several strategies can be implemented to assign the \ac{ccc}.
For instance, the \ac{ccc} is assigned by the \ac{ce} or determined based on internal ranks~\cite{Sidik2020LE}.

The \ac{ccc} is responsible for managing the prerequisites for the respective cycle, as well as coordinating the data collection and publication round.
To prepare the data collection, the \ac{ccc} initializes a \ac{cnt}.
The \ac{cnt} needs to implement the basic functionalities to \texttt{add}, \texttt{count}, and \texttt{remove} an element.
The \ac{cnt} is used for distributed counting between the \acp{gw} during the collection round and designation of publication responsibility in the publication round.
Once the \ac{cnt} is initialized, the \ac{ccc} forwards it to its successor and starts of the data collection round.

%%%%%%%%%%%%%%%%%%%%%
\subsubsection{Decentralized Collection Round}

Upon receiving \ac{cnt}, each \ac{gw} adds the counts of its data points to the \ac{cnt}, and forwards it to the succeeding \ac{gw} to continue the data collection process.
As in~\cite{Jha2023zAnon}, each \ac{gw} remembers values reported by their \acp{sn} and the number of its occurrences within $\Delta t$.
First, the \ac{gw} ensures that only measurements with timestamps within $\Delta t$ are considered while older data points are deleted (\texttt{apply\_delta\_t()}). 
Next, the \ac{gw} requests new measurements from its associated \acp{sn} and increases the respective counters in the \ac{cnt}.
Additionally, the \ac{gw} stores the new records containing measured values and the corresponding \ac{sn} IDs for the duration of the clock cycle because the counted values from all \acp{gw} need to be exchanged first before potentially publishing data points. 
As long as the ring has not been completely traversed the current \ac{gw} forwards the updated \ac{cnt} to its successor, which then repeats the process.
When the \ac{ccc} receives the \ac{cnt}, the round has been completed, and all data points from \acp{gw} have been processed.

%%%%%%%%%%%%%%%%%%%%%
\subsubsection{Ensure minimum count (\zanon)}
Once the collection round is completed and \ac{ccc} has received the updated \ac{cnt}, it proceeds with the post-processing to prepare the data publication.
First, the \ac{ccc} reads \ac{cnt}, and ensures that only data points meeting the required minimum count~$z$ are considered for publication.
To this end, it subtracts $z-1$ from all the value counts, those counts lower than $z-1$ are set to zero.
This cleaned \ac{cnt} is then used for the publication round.

%%%%%%%%%%%%%%%%%%%%%
\subsubsection{Decentralized Publication Round}

In the publication round, the \ac{cnt} is used to exchange information among \acp{gw} about whether a value has been counted at least~$z$ times during the collection round.
The \ac{ccc} initiates the publication round by forwarding the cleaned \ac{cnt} to its successor.
Upon receiving the \ac{cnt}, each \ac{gw} can check if the counters in the \ac{cnt} are positive for the measurement values reported by its \acp{sn} during data collection.
This means the value occurred more than~$z$ times, and the \ac{gw} may publish it by sending a tuple with value, ID, and timestamp to the \ac{ce}.
It then decrements the counter in the \ac{cnt} by~1, ensuring no more occurrences are published than permitted.
Last, the updated \ac{cnt} is forwarded to the successor.

When publication proceeds in the order defined by the system's ring structure, a bias may arise, as the \acp{gw} immediately following the \ac{ccc} are more likely to publish.
To mitigate this bias, we introduce a publication probability~$p_{pub}$, which the \ac{ccc} sets.
The \acp{gw} publish a value with probability~$p_{pub}$, reducing the publication bias.
Since this step is probabilistic, not all values may be published after the round completes.
Therefore, the \ac{ccc} can initiate an additional round with $p_{pub}=1$ to guarantee publication of all remaining values.

If the bias is acceptable, publication may be enforced in the first round, thus, omitting additional publication rounds.
It is important to note that, in general, it is not necessary for the \ac{cnt} to be empty to stop the publication round, since all values that occurred within~$\Delta t$ are counted, however, only the data points of the current clock cycle are available for publication.
The entire procedure is repeated in the next clock cycle with a new designated \ac{ccc}.

%% file: 5_privInDez.tex
\section{Privacy-Preserving Aggregation}
\label{sec:privInDez}

\dez realizes \zanon in a distributed system by adding minimal coordination between \acp{gw}.
We consider a network with a set $\mathcal{S}$ of \acp{sn}, and a set $\mathcal{N}=\{N_1,\dots,N_n\}$ of \acp{gw} arranged in a ring, where $N_1$ is the \ac{ccc}, without loss of generality.
Each node~$N_i$ holds a local multiset~$\mathcal{M}_i \subseteq \mathcal{U}$ with measurement values of its associated $\{\ac{sn}_i\}$, where $\mathcal{U} \subseteq \mathbb{Q}$.

The basic functionality of \dez does not offer protection against an \ac{hbc}, since the adversary can observe counts for reported measurements during data collection.
To add privacy protection, we propose the combination of a \emph{stochastic} counting structure and add a secure sum strategy.
While cryptographic solutions could also provide privacy, we focus on lightweight mechanisms that remain viable for sensor networks with low-power entities and high measurement frequency.
In line with this design goal, the stochastic \ac{cnt} is a suitable choice, as it reduces the space complexity for message transmission between \acp{gw}.

\subsection{Stochastic Counting Structure}

Specifically, we choose a \ac{cbf}~\cite{Fan2000CBF,Gakhov2019CBF} to represent the aggregated multiset.
However, the data structure can be replaced with any solution that supports distributed counting and information exchange regarding the existence of values.
Given their memory and computing efficiency that seem reasonable for implementation in sensor networks, \acp{cbf} are a suitable option that we have chosen deliberately for our implementation of the protocol.

A \ac{cbf} is an array of counters to record the frequency of value occurrences with the help of independent hash functions.
More precisely, a \ac{cbf} is defined as a tuple~$\mathsf{CBF} = \langle m, k, b, H, C \rangle$, where $m \in \mathbb{N}$ denotes the length of the counter array~$C \in \mathbb{N}^m$, and $k \in \mathbb{N}$ denotes the number of hash functions~$H = \{h_1, \dots, h_k\}$ with $h_j:\mathcal{U} \to \{0,\dots,m-1\}$.
Each counter can represent up to $2^b$ occurrences, where $b$ is the number of allocated bits per counter.

To add an element~$x \in \mathcal{U}$, all $k$ hash functions are evaluated, and the corresponding counters are incremented by~1:
\begin{align}
  \texttt{CBFAdd:} \forall j \in \{1,\dots,k\}: \; C[h_j(x)] \gets C[h_j(x)] + 1
\end{align}

Deletion follows the same procedure as addition,
but decrements the corresponding counters by~1.
The frequency of~$x$ can then be estimated as:
\begin{align}
  \texttt{CBFCount:} \hat{f}(x) = \min_{j=1}^k C[h_j(x)].
\end{align}

Due to potential hash collisions, false positives are possible, meaning counts could be greater than~0 even if the element was never added.
For a given expected number of elements~$\mathbb{X}$, the parameters $m$ and $k$ can be adjusted according to the accepted false positive probability $P_{fp}$~\cite{Gakhov2019CBF}: 
\begin{align}\label{al:cbf_params}
    m = - \frac{\mathbb{X} \cdot ln(P_{fp})}{(ln 2)^2} \qquad \text{and} \qquad
    k = \frac{m}{\mathbb{X}} \cdot ln 2
\end{align}
The parameter estimation should be performed based on the specific use case, taking into consideration the required privacy and utility guarantees.

Implementing a stochastic data structure also offers the opportunity to enhance privacy, because only the number of value occurrences needs to be shared for \zanon and no additional details of the data points are required.
However, directly applying a \ac{cbf} carries the risk of reverse-engineering measurement values from the stored counts~\cite{Reviriego2023CBFPriv}.
To address this limitation, we introduce an additional layer of protection through secure summation.

%%% 
\subsection{Secure Summation}

We propose the use of a secure sum algorithm to ensure that intermediate nodes cannot infer the exact number of occurrences during data collection.
Our design builds upon the \ac{bss}~\cite{Clifton2002SSum,Ranbaduge2020SMC}, which can be seamlessly integrated in \dez.
Here, a perturbation masks partial sums during aggregation, ensuring that no intermediate node learns individual contributions~\cite{Tschorsch2013distCount}.
This is particularly important for values that occur fewer than~$z$ times and must remain private.

In our design, secure sum is achieved by having~$N_1$ add a random perturbation~$R \in \mathbb{N}^m$ to the initial counter array and removing it at the end.
In~\cite{Clifton2002SSum}, $R$ is uniformly chosen from $[1, \cdots , n]$.
Note that, with domain knowledge, the maximum required noise may be reduced, which can further reduce the counter size~$b$.
We accordingly adapt the initialization and publication preparation by having the \ac{ccc} add the perturbation to all bits of the \ac{cbf} during initialization and remove it again during publication preparation.

\paragraph*{Initialization} 
$N_1$ samples $R^m \in \mathbb{N}^m$, and initializes the \ac{cnt} with $C \gets$ \texttt{CBFAdd}$(\{R^m\})$, and forwards $C$ to $N_2$.

\paragraph*{Aggregation}
When $N_2$ receives $C$, the decentralized collection round is executed. 
At each $N_i$, with $2 \leq i \leq n$, the node updates $C$ by adding each element $x \in \mathcal{M}_i$ to the \ac{cnt}, i.e., $C \gets \texttt{CBFAdd}(x)$.
Finally, the updated \ac{cnt} is forwarded to the next node, with $N_n$ sending it back to $N_1$.
Due to the perturbation $R$, privacy protection against an \ac{hbc} is provided during data collection.

\paragraph*{Finalization} 
Before initiating the publication phase, $N_1$ receives the final \ac{cnt}~$C$ and removes the initial perturbation~$R$, yielding~$\langle m, k, b, H, C \rangle$, a \ac{cbf} representing the aggregated multiset $\mathcal{M}=\bigcup_{i=1}^n \mathcal{M}_i$.
For any $x \in \mathcal{U}$, the estimated frequency is given by $\hat{f}(x) = \texttt{CBFCount}(x)$.
Correctness of the distributed counts follows, since $R$ cancels out in this finalization step.

%%%%%%%%%%%%%%%%%%%%%%%%%%%%%%%%%%%%%%%%%%%%%
\subsection{ID Masking}

Even though the stochastic \ac{cnt} and secure sum offer protection during the publication process, knowledge can still be inferred based on the published data tuples.
In \zanon, published tuples comprise a timestamp, a measurement value, and a client ID.
To avoid a direct link between IDs and client ID,~\cite{Jha2023zAnon} assign rotating pseudonyms.
For \dez, this could be implemented by using rotating pseudonyms for the \acp{gw}.
Presumably, this comes at a cost to data utility; an implementation and detailed analysis are left for future work.

To further strengthen privacy, the IDs of individual \acp{sn} can be replaced with the ID of their coordinating \ac{gw}.
This masks client identities and eliminates the need for rotating \ac{sn} pseudonyms.
The approach is conceptually similar to Network Address Translation~\cite{rfc3022}, where multiple devices share a common identifier.
The approach changes linking and tracing possibilities between clock cycles.
However, the \ac{ce} can still infer that each \ac{sn} contributes exactly one measurement per cycle, so data utility is preserved.

%%%%%%%%%%%%%%%%%%%%%%%%%%%%%%%%%%%%%%%%%%%%%
%%%%%%%%%%%%%%%%%%%%%%%%%%%%%%%%%%%%%%%%%%%%%

\subsection{Privacy Analysis of \dez}
\label{subsec:privDez}

By combining a stochastic data structure with a lightweight secure summation strategy, \dez achieves local \zanon in a distributed network with minimal coordination.
It provides the same output privacy as centralized \zanon~\cite{Jha2023zAnon}, while drastically reducing the trust placed in the \ac{ce}.
In \dez, the \ac{ce} only learns information about data tuples that are published, since the anonymization process is only performed on the \acp{gw}.
Accordingly, the \ac{ce} does not receive any information about measurements that occur fewer than~$z$ times, unlike when the \ac{ce} applies \zanon centrally, which would allow the \ac{ce} to analyze information from all tuples.

Currently, the \ac{ce} can link messages and the sending \ac{gw} since it is aware of the message origin.
Potential mitigation techniques include publication via proxy, but the performance and privacy impact of this needs further analyses.
As \acp{sn} can be masked by their associated \acp{gw}, we expect our approach to provide sufficient privacy for most practical use cases.

When removing the perturbation $R$, the \ac{ccc} is the only \ac{gw} that is capable of identifying values that occurred fewer than~$z$ times,
We therefore propose to rotate the \ac{ccc} each clock cycle to reduce the capability of a single \ac{gw} to correlate values across multiple clock cycles.

Beyond the \ac{ccc}, each \ac{gw} only observes perturbed intermediate aggregates during secure sum and thus cannot distinguish genuine measurements from injected noise.
However, if two \acp{gw} collude and happen to be neighbors of an honest \ac{gw} in the ring, they can compare the incoming and outgoing values and thereby extract information about that \ac{gw}'s values.
That is, \dez building upon \ac{bss} does not provide full collusion resistance~\cite{Ranbaduge2020SMC}, but offers probabilistic protection:
in a pseudo-random ring, the probability of a successful targeted collusion attack is $p_{att} = \frac{k}{(n-1)} \cdot \frac{(k-1)}{(n-2)}$ for $k$ attackers among $n$ nodes.
Table~\ref{tab:coll_prob} shows this probability of success for different shares $\mathcal{K}$ of $k$ attackers among $n$ nodes.
For $n=100$ \acp{gw} and $k=20$ attackers, this results in $p_{att} = 0.04$.
Even for a comparatively high share of adversarial \acp{gw}, the estimated attack success probability remains low.
We thus consider \ac{bss} a reasonable choice for our setting.

\begin{table}[]
  \vspace{1em}
\caption{Collusion probability $p_{att}$ for malicious share $\mathcal{K}$}
\label{tab:coll_prob}
\centering
\begin{tabular}{c||l l l l l}
\toprule
\bfseries $\mathcal{K}$ & 0.1 & 0.2 & 0.3 & 0.4 & 0.5\\
\midrule
\bfseries $p_{att}$ & 0.009 & 0.04 & 0.09 & 0.16 & 0.25\\
\bottomrule
\end{tabular}
\end{table}

%% file: 6_methodology.tex
\section{Methodology}
\label{sec:methodology}

In the following, we outline our simulation methodology. 
We base our simulation on an exemplary smart metering use case and consider three main scenarios: centralized, fully decentralized, and \dez.

%%%%%%%%%%%%%%%%%%%%%
\paragraph*{Simulation Data}

To get a data basis, we synthetically generate energy consumption data for different client types based on publicly available standard load profiles.\footnote{\url{https://www.bdew.de/energie/standardlastprofile-strom/}}
These profiles include representative consumption patterns, e.g., for households, farms, and local businesses, each exhibiting different usage behaviors.
The profiles provide values for every 15~minutes and are normalized to a maximum annual consumption of 1,000~kWh/a.
Even tough we do not aim to realize \zanon for electricity metering in particular, we want to utilize realistic data.
We therefore scale the values to typical consumption ranges for each client type, using publicly available statistics from the Federal Statistic Office of Germany.\footnote{\url{https://www.destatis.de/DE/Themen/Gesellschaft-Umwelt/Umwelt/UGR/private-haushalte/Tabellen/stromverbrauch-haushalte.html}}
The references for the scaling values are documented in our GitHub project.
Using these profiles, we model an exemplary district in Berlin, Germany.
We first define a representative distribution of various client type profiles.
Then, for each simulation, the type of each \ac{sn} can be randomly determined based on the underlying profile distribution, and the measurement values are generated based on the corresponding standard load profile.

Note that the measurement data can take arbitrary floating-point values, which can distort the anonymization.
To address this, we discretize the measurements by rounding them to key values (i.e., buckets), which are then used to count occurrences.
In our simulation, the rounding level increases with larger values, under the assumption that larger measurements exhibit greater variability.

%%%%%%%%%%%%%%%%%%%%%
\paragraph*{Simulation Scenarios}
For our simulation, we implement three scenarios: centralized, fully decentralized, and \dez (i.e., decentralized with coordination).
The centralized and fully decentralized scenarios serve as reference points.
In the centralized setup, \zanon is applied at the \ac{ce}, while in the fully decentralized, each \ac{gw} applies local \zanon without coordination, i.e., on the locally available data only.

In all scenarios, the number of \acp{gw} and the maximum number of \acp{sn} connected to each \ac{gw} can be adjusted.
The number of \acp{sn} is drawn from a normal distribution, with the peak occurring at half of the maximum number.
We also evaluate other parameters, such as the value of~$z$ and the client types.
This enables us to examine how the network architecture influences the achievable level of anonymity.

We validated our centralized implementation against the original implementation of \zanon~\cite{Jha2023zAnon}.
Interested readers can find the results in Appendix~\ref{app:modelval}.

%% file: 7_evaluation.tex
\section{Evaluation}
\label{sec:eval}

 \pgfplotsset{
    tick label style={font=\footnotesize},
    label style={font=\footnotesize},
    legend style={font=\footnotesize},
}

\pgfplotsset{
    colormap name=viridis,
}

%%% colors for scenarios
\pgfplotscreateplotcyclelist{mycolor}{
    [indices of colormap=
    {0, 14, 6} of viridis]
}

\pgfplotscreateplotcyclelist{mylinestyles}{
    {solid,mark=*},
    {densely dashed,mark=10-pointed star},
    {dashed,mark=o},
    {dashdotted,mark=triangle*},
    {densely dotted,mark=diamond*},
    {dotted,mark=square*}
}

\newcommand{\plotheight}{1.5in}

\pgfplotsset{
    discard if not/.style 2 args={
        x filter/.code={
            \edef\tempa{\thisrow{#1}}
            \edef\tempb{#2}
            \ifx\tempa\tempb
            \else
                \def\pgfmathresult{inf}
            \fi
        }
    }
}

%%% color for n_gw rotation
\pgfplotscreateplotcyclelist{color_ngw}{
    [samples of colormap=
    {7 of viridis }
    ]
}

%%% publication ratio vs z value, depending on #GWs
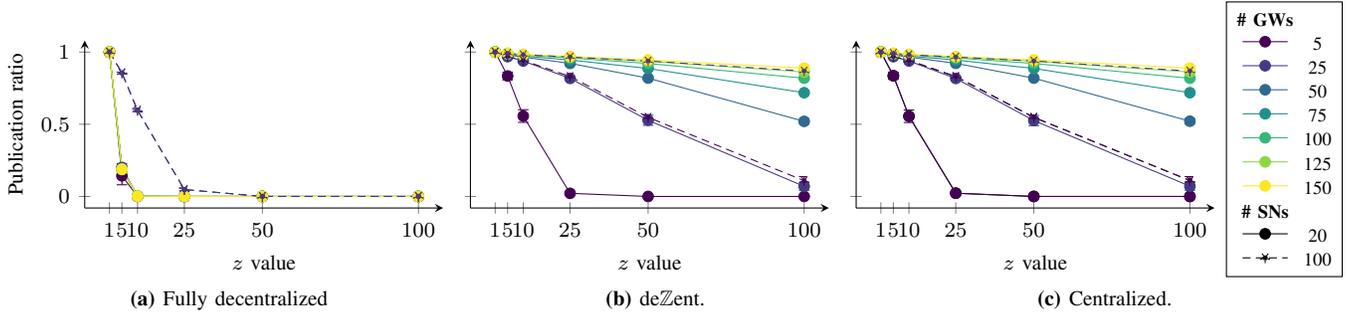
\begin{figure*}[tbhp]
    \centering
    \subfloat[Fully decentralized]{
        \hspace{-0.2cm}
        \input{pub_ratio_full_dec.tikz}
        \label{fig:pub_ratio_fd}
    }
    \subfloat[\dez.]{
        \hspace{-0.7cm}
        \input{pub_ratio_deZent.tikz}
        \label{fig:pub_ratio_deZ}
    }
    \subfloat[Centralized.]{
        \hspace{-0.7cm}
        \input{pub_ratio_centralized.tikz}
        \label{fig:pub_ratio_cent}
    }
    \caption{\footnotesize Average ratio of published tuples in different scenarios depending on value of $z$, 10 simulation runs.}
    \label{fig:pub_ratio}
    \vspace{-15pt}
\end{figure*}

%%% message count average across 10 runs 
\begin{figure*}[tbhp]
    \centering
    \subfloat[\ac{gw} $\leftrightarrow$ CE.]{
        \input{avg_msg_cnt_gwce_p_sce.tikz}
        \label{fig:msg_cnt_bar_gwce}
    }
    \subfloat[\ac{gw} $\leftrightarrow$ CE.]{
        \hspace{-0.3cm}
        \input{msgcnt_gwce_line_2sce.tikz}
        \label{fig:msg_cnt_line_gwce}
    }
    \subfloat[\ac{gw} $\leftrightarrow$ \ac{gw}/\ac{sn}.]{
        \hspace{-0.3cm}
        \input{avg_msg_cnt_gwgw_p_sce.tikz}
        \label{fig:msg_cnt_bar_gwgw}
    }
    \caption{\footnotesize Average number of messages transmitted during simulation, 10 simulation runs.}
    \label{fig:msg_cnt}
    \vspace{-15pt}
\end{figure*}
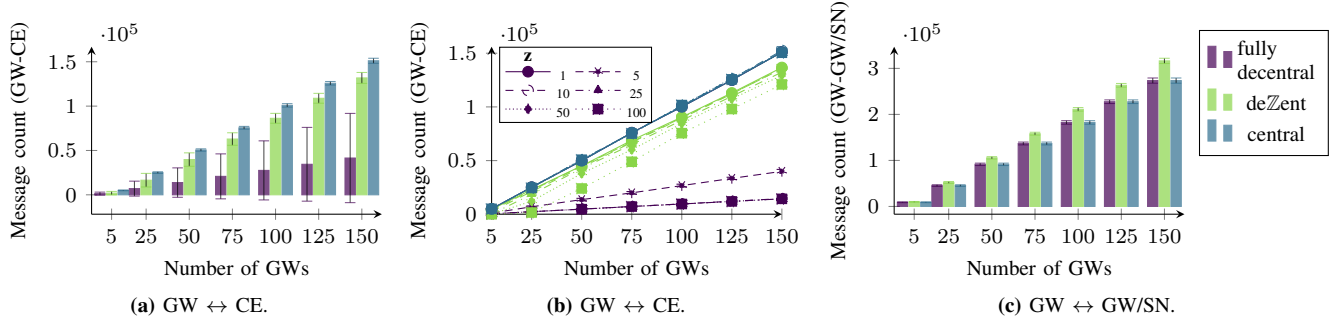

%%%%%%%%%%%%%%%%%%%%%%%%%%%%%%%%%%%%%%%%%%5

In the following, we present the results of our simulations and compare the three scenarios explained above, centralized, \dez, and fully decentralized.
We evaluate various metrics to estimate the costs and benefits caused by the decentralized implementation of \zanon with coordination.
We conducted the simulation for $z \in \{1, 5, 10, 25, 50, 100\}$ and the number of \acp{gw} in $\{5, 25, 50, 75, 100, 125, 150\}$.
We run the simulation 10 times for each set of parameters.
In the majority of experiments, we observe low standard deviation, displayed with error bars.

%%%%%%%%%%%%%%%%
\subsection{Results}

%%%%%%%%%%%%%%%%
\subsubsection{Publication Ratio}
To evaluate data utility in decentralized \zanon, we analyze the ratio of published tuples as a metric.
In Fig.~\ref{fig:pub_ratio}, we show the publication ratio depending on the value of~$z$ and the number of \acp{gw} in the system.

We observe that the publication ratio in the centralized scenario and in \dez is nearly identical, as depicted in Fig.~\ref{fig:pub_ratio_deZ} and~\ref{fig:pub_ratio_cent}.
This suggests that data utility for further analyses remains consistent across both scenarios.
Additionally, we notice that the publication ratio increases with a higher number of \acp{gw} in the system each with the same number of \acp{sn}, which results in a higher total number of entities.
This aligns with our expectations, as a larger number of \acp{sn} results in more value occurrences and, consequently, a greater number of data points that can be published.
In contrast to a fully decentralized scenario, the benefits of the scenarios using centralized \zanon or \dez become clear.
Fig.~\ref{fig:pub_ratio_fd} clearly shows that in a fully decentralized scenario the number of \acp{sn} connected to a \ac{gw} is the key factor influencing the publication ratio.
As the number of \acp{sn} increases, the publication ratio also rises (dashed line, star).

Furthermore, the total number of entities in the system has a greater impact on the results than their distribution between \acp{gw} and \acp{sn}.
This can be seen in Fig.~\ref{fig:pub_ratio_deZ}, where two different configurations with $\#\acp{gw} = 25, \#\acp{sn}=20$ (solid blue line) and $\#\acp{gw} = 5, \#\acp{sn}=100$ (dashed purple line) both result in approximately 250 entities (average $\#\acp{sn}$ corresponds to $\frac{\#\acp{sn}}{2}$), show similar outcomes.

To get an impression of the suitability of \dez for potential sensor networks, we also analyzed the publication ratio depending on hypothetical client types in a smart metering scenario.
Appendix~\ref{app:eval} shows the corresponding results.

%%%%%%%%%%%%%%%%
\subsubsection{Message Count}
To estimate the coordination overhead, we analyze the number of messages transmitted within the system.
The simulation results presented below were obtained with the maximum number of \acp{sn} set to 20.
Higher numbers of system entities, e.g., $\#\ac{sn} = 100$ were not feasible due to performance constraints.
Therefore, in this analysis, we vary only the number of \acp{gw} and the value of $z$.

In Fig.~\ref{fig:msg_cnt_bar_gwce}, we illustrate the number of messages transmitted between \acp{gw} and the \ac{ce}, with each color depicting the count for one scenario, respectively.
We observe that more messages are transmitted in the centralized scenario since all measured data points are forwarded to the \ac{ce}.
In both other scenarios only the anonymized data points are sent.

Fig.~\ref{fig:msg_cnt_line_gwce} shows the number of messages for different values of $z$. 
We note that in \dez more messages are sent towards the \ac{ce} than in the fully decentralized scenario.
This finding confirms that more data points can be published and forwarded to the \ac{ce} if coordination between the \acp{gw} is added to the local anonymization process.
This advantage becomes particularly apparent as the value of $z$ increases, leading to a decrease in the number of data points published.

Next, we examine the number of messages transmitted between the \acp{gw} themselves and between the \acp{gw} and the \acp{sn}.
Fig.~\ref{fig:msg_cnt_bar_gwgw} shows the message count for the three scenarios.
We observe that the message count is consistent in both the centralized and fully decentralized scenario.
Given that the system architecture is identical across all scenarios, this aligns with our expectations. 
Both scenarios transmit only the messages necessary to initiate the measurement process and those required to send the measurements from the \acp{sn} to the collecting \ac{gw}.
Furthermore, we observe that the message count in \dez is higher compared to the other scenarios.
This increase is anticipated due to the messages required for coordination among the~\acp{gw}.
However, this overhead corresponds approximately to that of the centralized scenario observed in Fig.~\ref{fig:msg_cnt_bar_gwce}.

%%%%%%%%%%%%%%%%
\subsubsection{Complexity}

In addition to the number of messages, the volume of transmitted data is decisive.
In \dez, the inter-\ac{gw} coordination transmits a \ac{cbf}.
Using the standard sizing from Eq.~\ref{al:cbf_params} with $P_{fp}=0.05$ and $\mathbb{X}=\num{1000}$ inserted items yields $m \approx 6235$ counters and an optimal $k \approx 4.32$ hash functions.
With 8-bit counters (counts up to 255), the filter size is $\approx 6.1$\,kB.
Thus, three passes around a ring of $|\mathcal{N}|$ gateways entail roughly $3\,|\mathcal{N}|\times 6.1$\,kB of coordination traffic per cycle.
By contrast, in the centralized scenario measurements from all \acp{sn} in $\mathcal{S}$ are sent from a \ac{gw} to the \ac{ce} with a payload in the order of tens of bytes (application data plus network headers).

Additionally, \acp{gw} need to store the measurements that associated \acp{sn} reported within $\Delta t$.
However, since \acp{gw} are system entities with more resources this seems like a reasonable assumption.

%%%%%%%%%%%%%%%%
\subsection{Discussion}

To provide privacy guarantees while minimizing trust, \dez requires lightweight coordination overhead.
The relative data volume of each scenario depends on $|\mathcal{N}|$, the number of decentrally anonymized tuples $|\mathcal{A}|$, and $|\mathcal{S}|$.
In typical deployments~$|\mathcal{N}|\ll |\mathcal{S}|$.
Especially for meaningful~$z$, when~$|\mathcal{A}|\ll |\mathcal{S}|$, the number of messages towards the \ac{ce} is reduced, which makes \dez competitive in total bytes despite a slight coordination overhead.

Both, the centralized scenario and \dez, achieve nearly identical publication rates and can provide high privacy protection, reflected by large values of~$z$.
In contrast, the fully decentralized scenario can only achieve lower values of~$z$ directly correlating to the number of \acp{sn} connected to each \ac{gw}.
Therefore, \dez presents a feasible solution to implement \zanon in a decentralized system by adding coordination.

As analyzed in Sec.~\ref{sec:privInDez} and shown by our results, \dez achieves the same privacy guarantees for its published data as the centralized implementation of \zanon.
While the exact order of tuples may differ, \dez produces an equivalent output, which is reflected in the number and ratio of published tuples.
Consequently, it achieves the same data utility and inherits the same privacy implications also regarding the probability of $k$-anonymous output as centralized \zanon~\cite{Jha2023zAnon}.

At the same time, \dez minimizes the need for trust towards a \ac{ce}.
Privacy risks and information leakage due to coordination between the \acp{gw} can be significantly reduced by carefully implementing noise for the \ac{cbf} and reasonable pseudonymization strategies to further protect the \acp{sn}.
In order to further protect \dez against more powerful attacks, future work might explore cryptographic extensions for coordination.
Since cryptography adds slight complexity, \dez might be more feasible for certain use cases due to its lightweight characteristics.
Depending on the exact use case and system capabilities, users could then choose between different versions of the protocol.

In summary, by implementing minimal coordination within the system, the \acp{gw} can protect the privacy of their clients in a distributed environment, assuming they adhere to an honest-but-curious attacker model toward other \acp{gw} and the \ac{ce}.

%% file: pub_ratio_full_dec.tikz
\begin{tikzpicture}

    \begin{axis}[
            height=\plotheight,
            width=0.35 \linewidth,
            axis y line=left,
            axis x line=bottom,
            enlargelimits=0.08,
            xlabel = {$z$ value},
            ylabel = {Publication ratio},
                error bars/y dir=both, 
                error bars/y explicit,
                error bars/error bar style={draw=black},
            xtick = data,
            x tick label style={/pgf/number format/1000 sep=},
            cycle multi list={
                mylinestyles \nextlist
                color_ngw
            }
        ]
        %%% fully dec
        \addplot+ [discard if not={n_gw}{5}] table[x=z, y=pub_ratio, y error = std] {avg_pub_ratio_fully_decentralized_nsm20.dat};

        \addplot+ [discard if not={n_gw}{25}] table[x=z, y=pub_ratio, y error = std] {avg_pub_ratio_fully_decentralized_nsm20.dat};

        \addplot+ [discard if not={n_gw}{50}] table[x=z, y=pub_ratio, y error = std] {avg_pub_ratio_fully_decentralized_nsm20.dat};

        \addplot+ [discard if not={n_gw}{75}] table[x=z, y=pub_ratio, y error = std] {avg_pub_ratio_fully_decentralized_nsm20.dat};

        \addplot+ [discard if not={n_gw}{100}] table[x=z, y=pub_ratio, y error = std] {avg_pub_ratio_fully_decentralized_nsm20.dat};

        \addplot+ [discard if not={n_gw}{125}] table[x=z, y=pub_ratio, y error = std] {avg_pub_ratio_fully_decentralized_nsm20.dat};

        \addplot+ [discard if not={n_gw}{150}] table[x=z, y=pub_ratio, y error = std] {avg_pub_ratio_fully_decentralized_nsm20.dat};

        \addplot+ [discard if not={n_gw}{5}] table[x=z, y=pub_ratio, y error = std] {avg_pub_ratio_fully_decentralized_nsm100.dat};

        \addplot+ [discard if not={n_gw}{25}] table[x=z, y=pub_ratio, y error = std] {avg_pub_ratio_fully_decentralized_nsm100.dat};

    \end{axis}

\end{tikzpicture}

%% file: pub_ratio_deZent.tikz
\begin{tikzpicture}

    \begin{axis}[
            height=\plotheight,
            width=0.35 \linewidth,
            axis y line=left,
            axis x line=bottom,
            enlargelimits=0.08,
            xlabel = {$z$ value},
            ylabel = \empty,
            yticklabel=\empty,
                error bars/y dir=both, 
                error bars/y explicit,
                error bars/error bar style={draw=black},
            xtick = data,
            x tick label style={/pgf/number format/1000 sep=},
            cycle multi list={
                mylinestyles \nextlist
                color_ngw
            }
        ]
        %%% fully dec
        \addplot+ [discard if not={n_gw}{5}] table[x=z, y=pub_ratio, y error = std] {avg_pub_ratio_deZent_nsm20.dat};

        \addplot+ [discard if not={n_gw}{25}] table[x=z, y=pub_ratio, y error = std] {avg_pub_ratio_deZent_nsm20.dat};

        \addplot+ [discard if not={n_gw}{50}] table[x=z, y=pub_ratio, y error = std] {avg_pub_ratio_deZent_nsm20.dat};

        \addplot+ [discard if not={n_gw}{75}] table[x=z, y=pub_ratio, y error = std] {avg_pub_ratio_deZent_nsm20.dat};

        \addplot+ [discard if not={n_gw}{100}] table[x=z, y=pub_ratio, y error = std] {avg_pub_ratio_deZent_nsm20.dat};

        \addplot+ [discard if not={n_gw}{125}] table[x=z, y=pub_ratio, y error = std] {avg_pub_ratio_deZent_nsm20.dat};

        \addplot+ [discard if not={n_gw}{150}] table[x=z, y=pub_ratio, y error = std] {avg_pub_ratio_deZent_nsm20.dat};

        \addplot+ [discard if not={n_gw}{5}] table[x=z, y=pub_ratio, y error = std] {avg_pub_ratio_deZent_nsm100.dat};

        \addplot+ [discard if not={n_gw}{25}] table[x=z, y=pub_ratio, y error = std] {avg_pub_ratio_deZent_nsm100.dat};

    \end{axis}

\end{tikzpicture}

%% file: pub_ratio_centralized.tikz
\begin{tikzpicture}
    
    \begin{axis}[
            hide axis,
            height=\plotheight,
            width=0.35 \linewidth,
            enlargelimits=0.08,
            cycle list name=mylinestyles,
        ]
        \addplot+ [discard if not={n_gw}{5}] table[x=z, y=pub_ratio] {avg_pub_ratio_centralized_nsm20.dat};\label{pgf:plot_sm20}

        \addplot+ [ discard if not={n_gw}{5}] table[x=z, y=pub_ratio] {avg_pub_ratio_centralized_nsm100.dat};\label{pgf:plot_sm100}
    \end{axis}

    \begin{axis}[
            height=\plotheight,
            width=0.35 \linewidth,
            axis y line=left,
            axis x line=bottom,
            enlargelimits=0.08,
            xlabel = {$z$ value},
            ylabel = \empty,
            yticklabel=\empty,
                error bars/y dir=both, 
                error bars/y explicit,
                error bars/error bar style={draw=black},
            xtick = data,
            x tick label style={/pgf/number format/1000 sep=},
            cycle multi list={
                mylinestyles \nextlist
                color_ngw
            }
        ]
        %%% fully dec
        \addplot+ [discard if not={n_gw}{5}] table[x=z, y=pub_ratio, y error = std] {avg_pub_ratio_centralized_nsm20.dat};\label{pgf:plot_gw5}

        \addplot+ [discard if not={n_gw}{25}] table[x=z, y=pub_ratio, y error = std] {avg_pub_ratio_centralized_nsm20.dat};\label{pgf:plot_gw25}

        \addplot+ [discard if not={n_gw}{50}] table[x=z, y=pub_ratio, y error = std] {avg_pub_ratio_centralized_nsm20.dat};\label{pgf:plot_gw50}

        \addplot+ [discard if not={n_gw}{75}] table[x=z, y=pub_ratio, y error = std] {avg_pub_ratio_centralized_nsm20.dat};\label{pgf:plot_gw75}

        \addplot+ [discard if not={n_gw}{100}] table[x=z, y=pub_ratio, y error = std] {avg_pub_ratio_centralized_nsm20.dat};\label{pgf:plot_gw100}

        \addplot+ [discard if not={n_gw}{125}] table[x=z, y=pub_ratio, y error = std] {avg_pub_ratio_centralized_nsm20.dat};\label{pgf:plot_gw125}

        \addplot+ [discard if not={n_gw}{150}] table[x=z, y=pub_ratio, y error = std] {avg_pub_ratio_centralized_nsm20.dat};\label{pgf:plot_gw150}

        \addplot+ [discard if not={n_gw}{5}] table[x=z, y=pub_ratio, y error = std] {avg_pub_ratio_centralized_nsm100.dat};

        \addplot+ [discard if not={n_gw}{25}] table[x=z, y=pub_ratio, y error = std] {avg_pub_ratio_centralized_nsm100.dat};

    \end{axis}

    % Legende
    \matrix[
        matrix of nodes,
        anchor=west,
        draw,% Rahmen um Legende
        inner sep= 2,
    ]
    at([xshift=0.15 cm, yshift=-0.2cm]current axis.east){
        \scriptsize \textbf{\# GWs}\\
        \ref{pgf:plot_gw5}& \scriptsize 5\\
        \ref{pgf:plot_gw25}& \scriptsize 25\\
        \ref{pgf:plot_gw50}& \scriptsize 50\\
        \ref{pgf:plot_gw75}& \scriptsize 75\\
        \ref{pgf:plot_gw100}& \scriptsize 100\\
        \ref{pgf:plot_gw125}& \scriptsize 125\\
        \ref{pgf:plot_gw150}& \scriptsize 150\\
        \scriptsize \textbf{\# SNs}\\
        \ref{pgf:plot_sm20} & \scriptsize 20\\
        \ref{pgf:plot_sm100}& \scriptsize 100\\
    };
    
\end{tikzpicture}

%% file: avg_msg_cnt_gwce_p_sce.tikz
\begin{tikzpicture}
    \begin{axis}[
            ybar=.5pt,
                error bars/y dir=both, 
                error bars/y explicit,
                error bars/error bar style={draw=black},
            x tick label style={/pgf/number format/1000 sep=},
            xlabel = {Number of \acp{gw}},
            ylabel = {Message count (GW-CE)},
            axis y line=left,
            axis x line=bottom,
            height=\plotheight,
            width=0.3 \linewidth,
            %ybar interval=1,
            enlargelimits=0.08,
            bar width= 6,
            xtick = data,
            cycle list name=mycolor,
        ]
        \addplot+ [fill, opacity=0.7, discard if not={scenario}{fully_decentralized}] table[x=n_gw, y=msg_cnt_gw_ce, y error = std] {avg_msg_cnt_gw_ce.dat};\label{pgf:fd_bar}

    	\addplot+ [fill, opacity=0.7, discard if not={scenario}{deZent}] table[x=n_gw, y=msg_cnt_gw_ce, y error = std] {avg_msg_cnt_gw_ce.dat};\label{pgf:deZ_bar}

        \addplot+ [fill, opacity=0.7, discard if not={scenario}{centralized}] table[x=n_gw, y=msg_cnt_gw_ce, y error = std] {avg_msg_cnt_gw_ce.dat};\label{pgf:cent_bar}

    \end{axis}
\end{tikzpicture}

%% file: msgcnt_gwce_line_2sce.tikz
\begin{tikzpicture}

    \begin{axis}[
            height=\plotheight,
            width=0.3 \linewidth,
            enlargelimits=0.05,
            axis y line=left,
            axis x line=bottom,
            error bars/y dir=both, 
            error bars/y explicit,
            error bars/error bar style={draw=black},
            xlabel = {Number of \acp{gw}},
            ylabel = {Message count (GW-CE)},
            xtick = data,
            x tick label style={/pgf/number format/1000 sep=},
            legend style={at={(.98,0.95)},
                anchor=north east, legend columns=-1},
            cycle multi list={
                mycolor \nextlist
                mylinestyles
            }
        ]
        %%% fully dec
        \addplot+ [ discard if not={z}{1}] table[x=n_gw, y=msg_cnt_gw_ce, y error = std] {msg_cnt_gw_ce_fully_decentralizedavg_per_z.dat};\label{pgf:plot_z1}
        \addplot+ [ discard if not={z}{5}] table[x=n_gw, y=msg_cnt_gw_ce, y error = std] {msg_cnt_gw_ce_fully_decentralizedavg_per_z.dat};\label{pgf:plot_z5}
        \addplot+ [discard if not={z}{10}] table[x=n_gw, y=msg_cnt_gw_ce, y error = std] {msg_cnt_gw_ce_fully_decentralizedavg_per_z.dat};\label{pgf:plot_z10}
        \addplot+ [discard if not={z}{25}] table[x=n_gw, y=msg_cnt_gw_ce, y error = std] {msg_cnt_gw_ce_fully_decentralizedavg_per_z.dat};\label{pgf:plot_z25}
        \addplot+ [discard if not={z}{50}] table[x=n_gw, y=msg_cnt_gw_ce, y error = std] {msg_cnt_gw_ce_fully_decentralizedavg_per_z.dat};\label{pgf:plot_z50}
        \addplot+ [discard if not={z}{100}] table[x=n_gw, y=msg_cnt_gw_ce, y error = std] {msg_cnt_gw_ce_fully_decentralizedavg_per_z.dat};\label{pgf:plot_z100}

        %%% deZent
        \addplot+ [discard if not={z}{1}] table[x=n_gw, y=msg_cnt_gw_ce, y error = std] {msg_cnt_gw_ce_deZentavg_per_z.dat};
        \addplot+ [discard if not={z}{5}] table[x=n_gw, y=msg_cnt_gw_ce, y error = std] {msg_cnt_gw_ce_deZentavg_per_z.dat};
        \addplot+ [discard if not={z}{10}] table[x=n_gw, y=msg_cnt_gw_ce, y error = std] {msg_cnt_gw_ce_deZentavg_per_z.dat};
        \addplot+ [discard if not={z}{25}] table[x=n_gw, y=msg_cnt_gw_ce, y error = std] {msg_cnt_gw_ce_deZentavg_per_z.dat};
        \addplot+ [discard if not={z}{50}] table[x=n_gw, y=msg_cnt_gw_ce, y error = std] {msg_cnt_gw_ce_deZentavg_per_z.dat};
        \addplot+ [discard if not={z}{100}] table[x=n_gw, y=msg_cnt_gw_ce, y error = std] {msg_cnt_gw_ce_deZentavg_per_z.dat};

        %%% central
        \addplot+ [discard if not={z}{1}] table[x=n_gw, y=msg_cnt_gw_ce, y error = std] {msg_cnt_gw_ce_centralizedavg_per_z.dat};
        \addplot+ [discard if not={z}{5}] table[x=n_gw, y=msg_cnt_gw_ce, y error = std] {msg_cnt_gw_ce_centralizedavg_per_z.dat};
        \addplot+ [discard if not={z}{10}] table[x=n_gw, y=msg_cnt_gw_ce, y error = std] {msg_cnt_gw_ce_centralizedavg_per_z.dat};
        \addplot+ [discard if not={z}{25}] table[x=n_gw, y=msg_cnt_gw_ce, y error = std] {msg_cnt_gw_ce_centralizedavg_per_z.dat};
        \addplot+ [discard if not={z}{50}] table[x=n_gw, y=msg_cnt_gw_ce, y error = std] {msg_cnt_gw_ce_centralizedavg_per_z.dat};
        \addplot+ [discard if not={z}{100}] table[x=n_gw, y=msg_cnt_gw_ce, y error = std] {msg_cnt_gw_ce_centralizedavg_per_z.dat};
        
    \end{axis}
    % Legende
    \matrix[
        matrix of nodes,
        anchor=north west,
        draw,% Rahmen um Legende
        inner sep=1,
    ]
    at([xshift=0.1cm, yshift=0cm]current axis.north west){
        \scriptsize \textbf{z}\\
        \ref{pgf:plot_z1}& \tiny 1 &
        \ref{pgf:plot_z5}& \tiny 5\\
        \ref{pgf:plot_z10}& \tiny 10 &
        \ref{pgf:plot_z25}& \tiny 25\\
        \ref{pgf:plot_z50}& \tiny 50 &
        \ref{pgf:plot_z100}& \tiny 100\\
    };
    
\end{tikzpicture}

%% file: avg_msg_cnt_gwgw_p_sce.tikz
\begin{tikzpicture}
    \begin{axis}[
            ybar=.5pt,
            error bars/y dir=both, 
            error bars/y explicit,
            error bars/error bar style={draw=black},
            x tick label style={/pgf/number format/1000 sep=},
            xlabel = {Number of \acp{gw}},
            ylabel = {Message count (GW-GW/SN)},
            axis y line=left,
            axis x line=bottom,
            height=\plotheight,
            width=0.3 \linewidth,
            enlargelimits=0.08,
            bar width= 6,
            xtick = data,
            cycle list name=mycolor,
            legend style={at={(1.05,1.1)},
                anchor=north west, legend columns=1,
                cells={align=left}},
        ]
        
        \addplot+ [ fill, opacity=0.7, discard if not={scenario}{fully_decentralized}] table[x=n_gw, y=msg_cnt_gw_gw, y error = std] {avg_msg_cnt_gw_gw.dat};

    	\addplot+ [fill, opacity=0.7, discard if not={scenario}{deZent}] table[x=n_gw, y=msg_cnt_gw_gw, y error = std] {avg_msg_cnt_gw_gw.dat};

        \addplot+ [fill, opacity=0.7, discard if not={scenario}{centralized}] table[x=n_gw, y=msg_cnt_gw_gw, y error = std] {avg_msg_cnt_gw_gw.dat};
        \legend{fully\\ decentral, \dez, central}
    \end{axis}
\end{tikzpicture}

%% file: 8_relatedWork.tex
\section{Related Work}
\label{sec:relWork}

Several approaches address the challenge of anonymizing continuous data streams, for instance by providing differential privacy~\cite{dwork2010DP,erlingsson2014Rappor}.
However, local differential privacy is not suitable for our use case and is more appropriate for aggregated statistics due to the noise it introduces, which affects utility.
Additionally, \kanon guarantees have been investigated for data streams, such as CASTLE~\cite{Cao2011Castle,Brunn2021} and its enhancements~\cite{Guo2013FAD,mohamed2019PrivDistDS,Yang2022Idea}.
\kanon has also been applied in Big Data contexts~\cite{Capitani2023kAnonBD,Shamsinejad2024}, however, focusing primarily on efficient anonymization via parallelization in centralized systems.
Most existing work addresses either continuous data streams~\cite{domingo2019AnonDS,Onesimu2021AnonIoT} or the distributed architectures~\cite{Tassa2013DistAn,Kohlmayer14DistAn}.
However, integrating both aspects is essential to advance anonymity in distributed sensor networks~\cite{Stegelmann2012GridPriv,Lee2015SecIot}.

Research on lightweight privacy-preserving distributed counting is particularly relevant to realize decentralization~\cite{Tschorsch2013distCount,Ashok2014Bloom,Voigt19DCount}.
Furthermore, multiparty computation and cryptography methods are frequently investigated as potential solutions~\cite{Goyal2022MPC,Lu2012Eppa,Ma2024MPC,Resende2021,Pillai2024MPC}.
Especially in the context of smart grid, this topic has gained attention.
\cite{rahman2017SMC-SG,mustafa2019SMC,khan2021fogSMC} propose solutions to integrate secure sum and multiparty computation for smart grid.
However, these approaches involve cryptography or computationally intensive processing that may not be feasible for resource-constrained sensor networks.

%% file: 9_conclusion.tex
\section{Conclusion}

In this paper, we analyzed the feasibility of implementing \zanon in a decentralized manner, enabling its application in distributed systems such as \ac{iot} networks.
To this end, we introduce \dez as a mechanism for local \zanon with minimal coordination among \acp{gw}.
This mechanism allows the deployment of \zanon independently of a \ac{ce}, relying instead on data exchange and coordination between \acp{gw}.
Our results indicate that the same level of privacy can be achieved with this distributed approach.
Furthermore, \dez provides enhanced privacy guarantees, as the \ac{ce} is no longer privy to all measurements without filtering.
Future research is required to determine the optimal parameter settings for specific use cases and to explore extensions that protect against stronger attackers.
In summary, \dez constitutes a lightweight solution for anonymization in sensor networks achieving privacy and utility comparable to the centralized solution while reducing trust requirements.

%% file: appendix.tex
\appendix
\section{Appendix}
\label{appendix}

%%%%%%%%%%%%%%%%%%%%%
\subsection{\dez -- Pseudocode}
\label{app:dezCode}

%%%%%%%%%%%%%%%%%%%%%
\begin{algorithm} [tbhp]
\caption{Local \zanon with coordination}
\label{alg:sys_setup}
    \centering
    \footnotesize
    \begin{algorithmic}[1]
        %%% define formatting
        \algrenewcommand{\algorithmicfunction}[1]{\textbf{def} #1}
        \algrenewcommand{\algorithmiccomment}[1]{%
                \scriptsize\ttfamily\textcolor{blue}{\hfill #1}
                \normalfont\footnotesize
            }
        \newcommand{\CommentBlock}[1]{%
                \Statex \Comment{\parbox[t]{\dimexpr\linewidth-2em}{\strut /* #1 */\vspace{2pt}}}%
                \normalfont\footnotesize
            }

        %%% define functions and shortcuts %%%
        \algdef{SxNi}[CLOCKCYCLE]{AtClockCycle}{EndAtClock}{2em}{\textbf{At each clock cycle:}}{}
        \algdef{SxNi}[CCC]{AtCCC}{EndAtCCC}{1.5em}{\textbf{At $\ac{ccc}$:}}{}
        \algdef{SxNi}[WITH]{With}{EndWith}{1.5em}{\textbf{With $p_{pub}$:}}{}

        %%% Algorithm %%%
        \Require Distributed network with semi-synchronized clocks in which \acp{gw} are interconnected in a ring architecture and collect measurements from their associated \acp{sn}
        \Ensure A joint $z$-anonymized data stream that can be published by the CE

        %%% MAIN function deZent %%%
        \AtClockCycle
            \State $t \gets$  current synchronized timestamp 
            \State $\ac{ccc} \gets $ designate a coordinating \ac{gw} \label{line:ccc_choice}
            \AtCCC
                \State $\ac{cnt} \gets$ \texttt{initialize\_cnt\_struct()} \label{line:init_cnt}
                \State \Call{CollectionRound}{$\ac{cnt},~t$} at $\ac{ccc}.successor$
                \State \Call{PrepareDataPublication}{$\ac{cnt},~z$}
                \State $p_{pub} \gets$ random initial publication probability
                \State \Call{PublicationRound}{$\ac{cnt},~p_{pub}, ~t$} at $\ac{ccc}.successor$
            \EndAtCCC
        \EndAtClock

    \vspace{10pt}
    %

        %%% COLLECTION ROUND %%%
        \Function{CollectionRound}{$\ac{cnt}, ~t$}
            \CommentBlock{delete counts that occurred longer than $\Delta$t ago but keep entries within time window}
            \State \texttt{apply\_delta\_t}(\ac{gw}$_i$.$cnt\_log, ~t$) \label{line:deltat}

            \CommentBlock{request and process new measurements}
            \State $new\_m \gets$ \ac{gw}$_i$.\{\ac{sn}$_i$\}.\texttt{get\_current\_measurement()}
            \State $\ac{gw}_i.current\_m \gets new\_m$ \label{line:storeM}
            \State $\ac{gw}_i.cnt\_log \gets$ \texttt{add}($new\_m$)
            \State $\ac{cnt} \gets$ \texttt{add}($new\_m$)

            \If{\ac{gw}$_i$ = \ac{ccc}} \Comment{// ring has been completed}
                \State \Return $\ac{cnt}$
            \Else
                \State \Call{CollectionRound}{$\ac{cnt}, ~t$} at $\ac{gw}_{i+1}$
            \EndIf
        \EndFunction

    \vspace{10pt}
    %

        %%% PREPARE PUBLICATION %%%
        \Function{PrepareDataPublication}{$\ac{cnt}, ~z$} \label{line:prep_pub}
            \CommentBlock{retain only value counts that are at least $z$}
            \State $\ac{cnt} \gets$ \texttt{get\_final\_cnt\_struct()} \label{line:post_cnt}
            \State \texttt{ensure\_minimum\_count}($\ac{cnt}, ~z$)
            \State \Return $\ac{cnt}$
        \EndFunction
    \vspace{10pt}
    %

        %%% PUBLICATION ROUND %%%
        \Function{PublicationRound}{\ac{cnt}, ~$p_{pub}$, ~t}
            \CommentBlock{publish $z$-anonymous tuple and update $\ac{cnt}$ accordingly}
            \State $\ac{cnt} \gets$ \Call{publish\_tuples}{$\ac{cnt},~p_{pub},~t$}
    
            \If{\ac{gw}$_i$ = \ac{ccc}} \Comment{// ring has been completed}
                
                \If{$\ac{cnt}$ is empty \textbf{or} $p_{pub} ==$ 1}
                    \State \Return
                \EndIf%}
                
            \EndIf
                \State \Call{PublicationRound}{$\ac{cnt}, ~p_{pub} \gets 1, ~t$} at $\ac{gw}_{i+1}$ 
        \EndFunction

    \vspace{10pt}
    %

        %%% PUBLISH TUPLES %%%
        \Function{publish\_tuples}{$\ac{cnt}, ~p_{pub}, ~t$}
            \ForAll{$(x, id) \in \ac{gw}_i.current\_m}$
                \If{$\ac{cnt}$.\texttt{count}($x$) $> 0$} \Comment{// x is $z$-anonymous} \label{line:zous_r}
                    \With \label{line:with_p}
                        \State send ($x$, $id$, $t$) to CE \Comment{// publish tuple}
                        \State $\ac{cnt}$.\texttt{remove}($m$) \label{line:rm_r}
                    \EndWith
                \EndIf
            \EndFor
            \State \Return $\ac{cnt}$
        \EndFunction

    \end{algorithmic}
\end{algorithm}

In the following, we show the pseudocode for the \dez algorithm as described in Se.~\ref{sec:deZent} and~\ref{sec:privInDez}.
The algorithm covers the main functionalities described above as well as implementation details for the realization of privacy guarantees, for instance, the implementation with $p_{pub}$~(line~\ref{line:with_p}).
The two functions to process \ac{cnt} (line~\ref{line:init_cnt} and~\ref{line:post_cnt}) refer to the privacy-preserving realization of \ac{cnt} such as adding (removing) noise for secure sum.
The exact steps of these functions depend on the chosen counting structure and required privacy guarantees.

%%%%%%%%%%%%%%%%%%%%%
%%%%%%%%%%%%%%%%%%%%%

\subsection{Model validation}
\label{app:modelval}

\pgfplotsset{
    tick label style={font=\footnotesize},
    label style={font=\footnotesize},
    legend style={font=\footnotesize},
} 

\pgfplotsset{
    colormap name=viridis,
} 

\pgfplotscreateplotcyclelist{centr_blue}{
    [indices of colormap= {6} of viridis]
}

\begin{figure}[tbhp]
    \centering
    \subfloat[Average difference of publication ratio, 10 runs.]{
        \input{ model_val_pubratio_diff.tikz}
        \label{fig:model_val_pr_diff}
    }
    \hfill
    \subfloat[Publication over time for z = 50, GW = 150, 1 run.]{
        \input{ model_val_npub_over_t.tikz}
        \label{fig:model_val_time}
    }
    \caption{\footnotesize Simulation results comparing the centralized scenario with our implementation and with that by~\cite{Jha2023zAnon}.}
    \label{fig:model_val}
\end{figure}
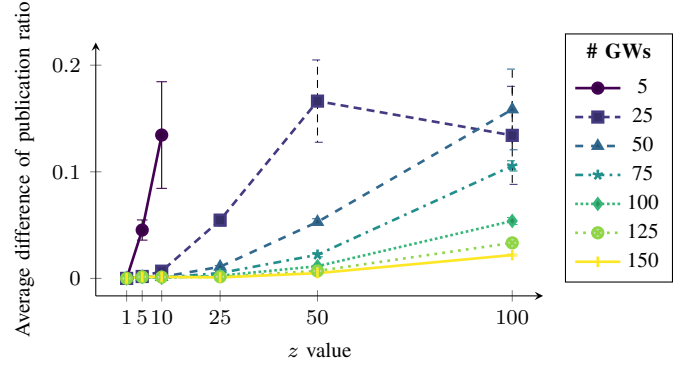
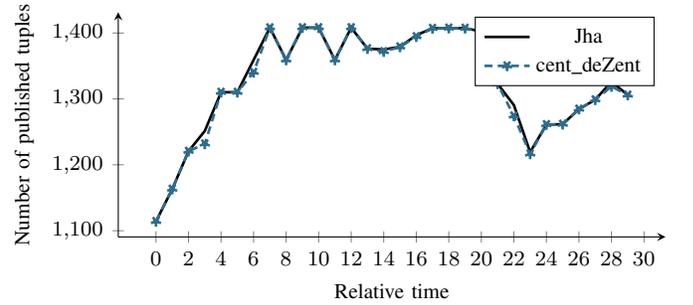

As mentioned above, we validated that our implementation meets \zanon guarantees that are comparable to the original implementation by~\cite{Jha2023zAnon}.
To this end, we compared the output generated by~\cite{Jha2023zAnon} with our simulation of a centralized scenario.

Due to implementation details, the exact tuples that are published might vary between the two implementations.
For instance, while \cite{Jha2023zAnon} processes incoming tuples sequentially, \dez first aggregates counts of all existing values in a distributed manner before publishing.
However, the overall trend remains consistent.
For the trivial case $z=1$, we confirmed that all measurement values were published in both implementations.

Fig.~\ref{fig:model_val_pr_diff} presents the difference in publication ratios between the two implementations. 
Both implementations produce nearly identical numbers of published measurements with only minor differences due to boundary effects of the implementation of window~$\Delta t$.
The difference is largest in scenarios where only a few tuples can be published, particularly for larger~$z$ values compared to smaller networks, for example, $z=50$ for 25 \acp{gw} with an average of 10 \acp{sn}.
In such scenarios, boundary effects appear to accumulate.
However, for larger networks, the difference remains minimal.
Apart from that, we verified that not only did the publication ratios align, but the same number of tuples were published for each measurement bucket.
The results are reproducible using the code available in our GitHub repository.

We also verified that the timing of the publication history aligns between the two implementations. 
Fig.~\ref{fig:model_val_time} depicts the corresponding data.
Further validations consistently confirmed comparable results, indicating that both published nearly identical data points.
Additional analyses are available in our GitHub.
Overall, this supports the conclusion that we effectively realize \zanon.

%%% color for client type rotation
\pgfplotscreateplotcyclelist{color_type}{
    [samples of colormap=
    {9 of viridis } ]
}

%% publication ratio per client type
\begin{figure*}[!t]
    \centering
    \subfloat[\# \ac{gw} = 5]{
        \input{ avg_smtype_pubrat_ngw5.tikz}
        \label{fig:pr_type_ngw5}
    }
    \subfloat[\# \ac{gw} = 75]{
        \hspace{-0.8cm}
        \input{ avg_smtype_pubrat_ngw75.tikz}
        \label{fig:pr_type_ngw75}
    }
    \subfloat[\# \ac{gw} = 150]{
        \hspace{-0.8cm}
        \input{ avg_smtype_pubrat_ngw150.tikz}
        \label{fig:pr_type_ngw150}
    }
    \caption{\footnotesize Ratio of published tuples in \dez depending on client type, 10 simulation runs.}
    \label{fig:smtype_pub_ratio}
    \vspace{0pt}
\end{figure*}
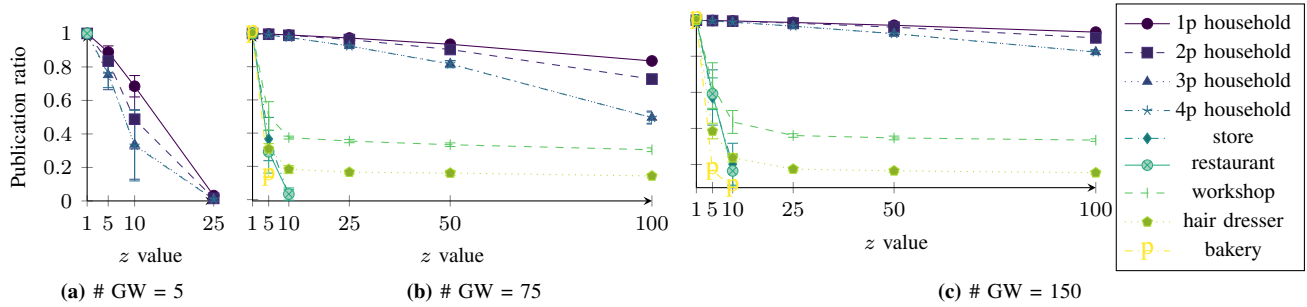

%%%%%%%%%%%%%%%%%%%%%
%%%%%%%%%%%%%%%%%%%%%

\balance

\subsection{Further Evaluation Results}
\label{app:eval}

To get an impression of the suitability of \dez for potential sensor networks, we additionally analyzed the publication ratio depending on hypothetical client types in a smart metering scenario.
This allows us to estimate wether \dez could be applied in scenarios with system entities that exhibit very different behaviors as it is the case in many sensor networks.
Fig.~\ref{fig:smtype_pub_ratio} shows the results for three simulations in which we set the number of \acp{gw} to $5$, $75$, and $150$, and the maximum number of \acp{sn} to $20$.
As anticipated, the publication ratio for a small network with 5 \acp{gw} decreases significantly even at small values of $z$, as shown in Fig.~\ref{fig:pr_type_ngw5}.
In addition, the standard deviation is highest in this scenario, presumably, because the ratio of published tuples varies significantly and depends on the exact network configuration.
In comparison, Fig.~\ref{fig:pr_type_ngw75} and~\ref{fig:pr_type_ngw150} show that in a larger network, data points can be published for nearly all client types.
The trend associated with increasing~$z$ is particularly noteworthy.
For instance, the publication ratio of workshops (see dashed turquoise line) does not decline further after an initial sharp drop.
Given the profile distribution indicating a scarcity of workshops in the network ($< 0.01\%$), this suggests that workshop data points can be published since they share consumption values with other entities in the system.
The consumption values for workshops resemble those of 3- and 4-person households, indicating a mutual dependency that facilitates the publication of the respective values.
Meanwhile, measurement values for client types with unusual measurements are not published as~$z$ increases.
To take advantage of this situation, one could consider using \zanon as a private filter.

%% file: model_val_pubratio_diff.tikz
\begin{tikzpicture}

    \begin{axis}[
            height=1.3*\plotheight,
            width=0.85*\linewidth,
            axis y line=left,
            axis x line=bottom,
            enlargelimits=0.08,
            xlabel = {$z$ value},
            ylabel = {Average difference of publication ratio},
                error bars/y dir=both, 
                error bars/y explicit,
                error bars/error bar style={draw=black},
            xtick = {1,5,10,25,50,100},
            x tick label style={/pgf/number format/1000 sep=},
            cycle multiindex* list={
                mylinestyles \nextlist
                color_ngw\nextlist
                mark list\nextlist
            },
            every axis plot/.append style={line width=1pt},
            legend style={at={(1.05,0.9)},
                anchor=north west, legend columns=1, name = legend, draw=none}
        ]
        %%% fully dec
        \addplot+ [discard if not={n_gw}{5}] table[x=z, y=pub_rat_diff, y error = std] {avg_pub_ratio_diff.dat};        

        \addplot+ [discard if not={n_gw}{25}] table[x=z, y=pub_rat_diff, y error = std] {avg_pub_ratio_diff.dat};

        \addplot+ [discard if not={n_gw}{50}] table[x=z, y=pub_rat_diff, y error = std] {avg_pub_ratio_diff.dat};

        \addplot+ [discard if not={n_gw}{75}] table[x=z, y=pub_rat_diff, y error = std] {avg_pub_ratio_diff.dat};

        \addplot+ [discard if not={n_gw}{100}] table[x=z, y=pub_rat_diff, y error = std] {avg_pub_ratio_diff.dat};

        \addplot+ [discard if not={n_gw}{125}] table[x=z, y=pub_rat_diff, y error = std] {avg_pub_ratio_diff.dat};

        \addplot+ [discard if not={n_gw}{150}] table[x=z, y=pub_rat_diff, y error = std] {avg_pub_ratio_diff.dat};

        \legend{5, 25, 50, 75, 100, 125, 150}
        
    \end{axis}

    \node [above,font=\bfseries] (legendtitle) at (legend.north) {\footnotesize \# \acp{gw}};
    \node [fit=(legendtitle)(legend),draw,inner sep=0pt] {};

\end{tikzpicture}

%% file: model_val_npub_over_t.tikz
\begin{tikzpicture}

    \begin{axis}[
            height=1.2*\plotheight,
            width=\linewidth,
            axis y line=left,
            axis x line=bottom,
            enlargelimits=0.08,
            xlabel = {Relative time},
            ylabel = {Number of published tuples},
            xtick = {0,2,...,30},
            x tick label style={/pgf/number format/1000 sep=},
            cycle multi list={
                mylinestyles \nextlist
                centr_blue},
            every axis plot/.append style={line width=1pt}, %change style of lines without affecting axis style
        ]
        %%% Jha
        \addplot [] table[x=time, y=tuple_count] {avg_jha_npub_over_time_cent_w_comm_zanon_z_50_dt_7260_nGw_150_distSm_normal_maxSm_20_seed_10.dat};

        %%% deZent
        \addplot+ [] table[x=time, y=tuple_count] {avg_deZ_npub_over_time_cent_w_comm_zanon_z_50_dt_7260_nGw_150_distSm_normal_maxSm_20_seed_10.dat};

        \legend{Jha, cent\_deZent}
        
    \end{axis}

\end{tikzpicture}

%% file: avg_smtype_pubrat_ngw5.tikz
% l_styles = ["1p_household", "2p_household", "3p_household", "4p_household", "store", "restaurant", "workshop", "hair_dresser", "bakery"]

\begin{tikzpicture}
    
    \begin{axis}[
            height=\plotheight,
            width= 0.18 \linewidth,
            enlargelimits=0.08,
            axis y line=left,
            axis x line=bottom,
            xlabel = {$z$ value},
            ylabel = {Publication ratio},
                error bars/y dir=both, 
                error bars/y explicit,
                error bars/error bar style={solid},
            xtick = data,
            x tick label style={/pgf/number format/1000 sep=},
            cycle multiindex* list={
                linestyles* \nextlist
                color_type\nextlist
                mark list\nextlist
            },
        ]
        %%% fully dec
        \addplot+ [discard if not={type}{1p_household}] table[x=z, y=pub_ratio, y error = std] {avg_smtype_pub_ratio_ngw5.dat};
        \addplot+ [discard if not={type}{2p_household}] table[x=z, y=pub_ratio, y error = std] {avg_smtype_pub_ratio_ngw5.dat};
        \addplot+ [discard if not={type}{3p_household}] table[x=z, y=pub_ratio, y error = std] {avg_smtype_pub_ratio_ngw5.dat};
        \addplot+ [discard if not={type}{4p_household}] table[x=z, y=pub_ratio, y error = std] {avg_smtype_pub_ratio_ngw5.dat};
        \addplot+ [discard if not={type}{store}] table[x=z, y=pub_ratio, y error = std] {avg_smtype_pub_ratio_ngw5.dat};
        \addplot+ [discard if not={type}{restaurant}] table[x=z, y=pub_ratio, y error = std] {avg_smtype_pub_ratio_ngw5.dat};
        \addplot+ [discard if not={type}{workshop}] table[x=z, y=pub_ratio, y error = std] {avg_smtype_pub_ratio_ngw5.dat};
        \addplot+ [discard if not={type}{hair_dresser}] table[x=z, y=pub_ratio, y error = std] {avg_smtype_pub_ratio_ngw5.dat};
        \addplot+ [discard if not={type}{bakery}] table[x=z, y=pub_ratio, y error = std] {avg_smtype_pub_ratio_ngw5.dat};

    \end{axis}

\end{tikzpicture}

%% file: avg_smtype_pubrat_ngw75.tikz
% l_styles = ["1p_household", "2p_household", "3p_household", "4p_household", "store", "restaurant", "workshop", "hair_dresser", "bakery"]

\begin{tikzpicture}
    
    \begin{axis}[
            height=\plotheight,
            width= 0.38 \linewidth,
            enlargelimits=0.08,
            axis y line=left,
            axis x line=bottom,
            xlabel = {$z$ value},
            ylabel = \empty,
            yticklabel=\empty,
                error bars/y dir=both, 
                error bars/y explicit,
                error bars/error bar style={solid},
            xtick = data,
            x tick label style={/pgf/number format/1000 sep=},
            cycle multiindex* list={
                linestyles* \nextlist
                color_type\nextlist
                mark list\nextlist
            },
        ]
        %%% fully dec
        \addplot+ [discard if not={type}{1p_household}] table[x=z, y=pub_ratio, y error = std] {avg_smtype_pub_ratio_ngw75.dat};
        \addplot+ [discard if not={type}{2p_household}] table[x=z, y=pub_ratio, y error = std] {avg_smtype_pub_ratio_ngw75.dat};
        \addplot+ [discard if not={type}{3p_household}] table[x=z, y=pub_ratio, y error = std] {avg_smtype_pub_ratio_ngw75.dat};
        \addplot+ [discard if not={type}{4p_household}] table[x=z, y=pub_ratio, y error = std] {avg_smtype_pub_ratio_ngw75.dat};
        \addplot+ [discard if not={type}{store}] table[x=z, y=pub_ratio, y error = std] {avg_smtype_pub_ratio_ngw75.dat};
        \addplot+ [discard if not={type}{restaurant}] table[x=z, y=pub_ratio, y error = std] {avg_smtype_pub_ratio_ngw75.dat};
        \addplot+ [discard if not={type}{workshop}] table[x=z, y=pub_ratio, y error = std] {avg_smtype_pub_ratio_ngw75.dat};
        \addplot+ [discard if not={type}{hair_dresser}] table[x=z, y=pub_ratio, y error = std] {avg_smtype_pub_ratio_ngw75.dat};
        \addplot+ [discard if not={type}{bakery}] table[x=z, y=pub_ratio, y error = std] {avg_smtype_pub_ratio_ngw75.dat};

    \end{axis}

\end{tikzpicture}

%% file: avg_smtype_pubrat_ngw150.tikz
% l_styles = ["1p_household", "2p_household", "3p_household", "4p_household", "store", "restaurant", "workshop", "hair_dresser", "bakery"]

\begin{tikzpicture}
    
    \begin{axis}[
            height=\plotheight,
            width= 0.38 \linewidth,
            enlargelimits=0.08,
            axis y line=left,
            axis x line=bottom,
            xlabel = {$z$ value},
            ylabel = \empty,
            yticklabel=\empty,
                error bars/y dir=both, 
                error bars/y explicit,
                error bars/error bar style={solid},
            xtick = data,
            x tick label style={/pgf/number format/1000 sep=},
            cycle multiindex* list={
                linestyles* \nextlist
                color_type\nextlist
                mark list\nextlist
            },
            legend style={at={(1.05,1.1)},
                anchor=north west, legend columns=1},
        ]
        %%% fully dec
        \addplot+ [discard if not={type}{1p_household}] table[x=z, y=pub_ratio, y error = std] {avg_smtype_pub_ratio_ngw150.dat};\label{pgf:1p_house}
        \addplot+ [discard if not={type}{2p_household}] table[x=z, y=pub_ratio, y error = std] {avg_smtype_pub_ratio_ngw150.dat};\label{pgf:2p_house}
        \addplot+ [discard if not={type}{3p_household}] table[x=z, y=pub_ratio, y error = std] {avg_smtype_pub_ratio_ngw150.dat};\label{pgf:3p_house}
        \addplot+ [discard if not={type}{4p_household}] table[x=z, y=pub_ratio, y error = std] {avg_smtype_pub_ratio_ngw150.dat};\label{pgf:4p_house}
        \addplot+ [discard if not={type}{store}] table[x=z, y=pub_ratio, y error = std] {avg_smtype_pub_ratio_ngw150.dat};\label{pgf:store}
        \addplot+ [discard if not={type}{restaurant}] table[x=z, y=pub_ratio, y error = std] {avg_smtype_pub_ratio_ngw150.dat};\label{pgf:restaurant}
        \addplot+ [discard if not={type}{workshop}] table[x=z, y=pub_ratio, y error = std] {avg_smtype_pub_ratio_ngw150.dat};\label{pgf:workshop}
        \addplot+ [discard if not={type}{hair_dresser}] table[x=z, y=pub_ratio, y error = std] {avg_smtype_pub_ratio_ngw150.dat};\label{pgf:hair}
        \addplot+ [discard if not={type}{bakery}] table[x=z, y=pub_ratio, y error = std] {avg_smtype_pub_ratio_ngw150.dat};\label{pgf:baker}
        \legend{1p household, 2p household, 3p household, 4p household, store, restaurant, workshop, hair dresser, bakery}
    \end{axis}

\end{tikzpicture}